# Circularly polarized metamaterial cage for homogeneous signal-to-noise ratio enhancement in magnetic resonance imaging

Yuhan Liu[1,2,†], Xia Zhu[1,2,†], Ke Wu[1,2], Stephan W. Anderson[2,3*], and Xin Zhang[1,2,4,5,6*]

[1]Department of Mechanical Engineering, Boston University, Boston, MA 02215, United States
[2]Photonics Center, Boston University, Boston, MA 02215, United States
[3]Chobanian & Avedisian School of Medicine, Boston University Medical Campus, Boston, MA, 02118, United States
[4]Department of Electrical & Computer Engineering, Boston, MA 02215, United States
[5]Department of Biomedical Engineering, Boston, MA 02215, United States
[6]Division of Materials Science & Engineering, Boston, MA 02215, United States

[†]These authors contributed equally to this work
*Corresponding author: Xin Zhang (xinz@bu.edu); Stephan W. Anderson (sande@bu.edu);
Contributing authors: yuhanyh@bu.edu; xiaz@bu.edu; wk0305ok@bu.edu

**ABSTRACT**
The signal-to-noise ratio (SNR) in magnetic resonance imaging (MRI) governs the quality of signal detection and directly impacts the clarity and reliability of the acquired images. Recent advances in metamaterials have enabled lightweight solutions with selective magnetic responses, offering a route to locally boost SNR in targeted anatomical regions but often with compromised field homogeneity. Here, a wireless metamaterial cage constructed from coaxial cables is engineered for homogeneous SNR enhancement at 3.0 T. With its cylindrical geometry and electromagnetic architecture, the device supports circularly polarized resonance through engineered phase-shifted currents, enabling selective and omnidirectional interaction with the rotating $B_1^-$ field to achieve uniform magnetic field distribution. Integrated with the body coil, the device yields a 32-fold SNR enhancement while maintaining comparable homogeneity to the body coil alone, exhibiting only 12.07% variation within the region of interest (ROI). Benchmarking against a state-of-the-art 16-channel extremity coil further shows that the metacage achieves at least 1.94-fold and 2.24-fold higher SNR in axial and coronal planes, respectively, and exhibits substantially lower SNR variation (12.07% compared to 54.83% for the extremity coil). The results establish the metacage as a compelling platform for next-generation wireless MRI technologies.

# 1 INTRODUCTION

Magnetic resonance imaging (MRI) enables non-invasive, high-resolution visualization of internal anatomical structures, establishing it as a pivotal modality in medicine [1]. Compared to other diagnostic imaging techniques, the non-ionizing and non-radioactive nature of MRI allows for safer longitudinal monitoring, therapeutic evaluation, and intervention planning, particularly in vulnerable populations [2-4]. The diagnostic power of MRI fundamentally relies on detecting the MRI signal, the weak radio frequency (RF) magnetic field $B_1^-$, produced by hydrogen nuclei in the human body [5,6]. These nuclei are initially aligned by a strong static magnetic field $B_0$ of the main magnet and subsequently excited by an RF transmit field $B_1^+$, typically generated using a birdcage coil (BC) [7,8]. This signal encodes critical information about anatomical structures and tissue properties, providing crucial insights into health and disease [9].

The static field B0 defines the operating field strength of the scanner (e.g., 1.5 T, 3.0 T, 7.0T) and determines the Larmor frequency of nuclear spins. Although continued advances in B0 stability and B1+ uniformity remain essential, particularly for higher field strengths where transmit inhomogeneity becomes more pronounced, improving the sensitivity and spatial homogeneity of the receive field B1- represents a practical and effective route for enhancing imaging quality across diverse MRI systems. In clinical practice, multi-channel coils are employed as receive-only arrays that operate alongside the BC for RF reception [14-16]. These localized receiver arrays are designed to maximize $B_1^-$ sensitivity within a specific region of interest (ROI), thereby enhancing the signal-to-noise ratio (SNR) and overall image quality. However, despite their diagnostic effectiveness, such coils are often bulky, rigid, and expensive [17-19]. Their fixed geometries can make it challenging to accommodate patients with varying body types, complicating positioning, reducing patient comfort, and prolonging setup time. Moreover, in phased-array MRI coil, each channel element detects signal from a limited spatial region, producing strong sensitivity close to the conductor and weaker sensitivity farther away. With only one or a few channels, this inherently leads to nonuniform signal distribution. In modern clinical practice, a common solution is to add the number of channels and strategically positioning them around the anatomy, their individual sensitivity patterns overlap and complement one another, effectively "shimming" the field to achieve a more uniform combined response in the ROI [20-22]. However, these improvements predominantly benefit regions close to the array elements, while the central region remains fundamentally constrained by the physics of ultimate intrinsic SNR (uISNR), making deep-region uniformity inherently more difficult to improve. Additionally, each receive channel requires its own preamplifier, detuning circuit, and decoupling network, leading to rapidly increasing system complexity and cost [23]. As a result, even advanced clinical scanners typically employ only 8-32 channels rather than hundreds. This fundamental trade-off between achievable field homogeneity and hardware feasibility has motivated the exploration of wireless metamaterial paradigms, which aim to deliver meaningful SNR enhancement and improved volumetric homogeneity in a lightweight, anatomically adaptable, and low-cost architecture enabled by advances in structural and materials design [24-28].

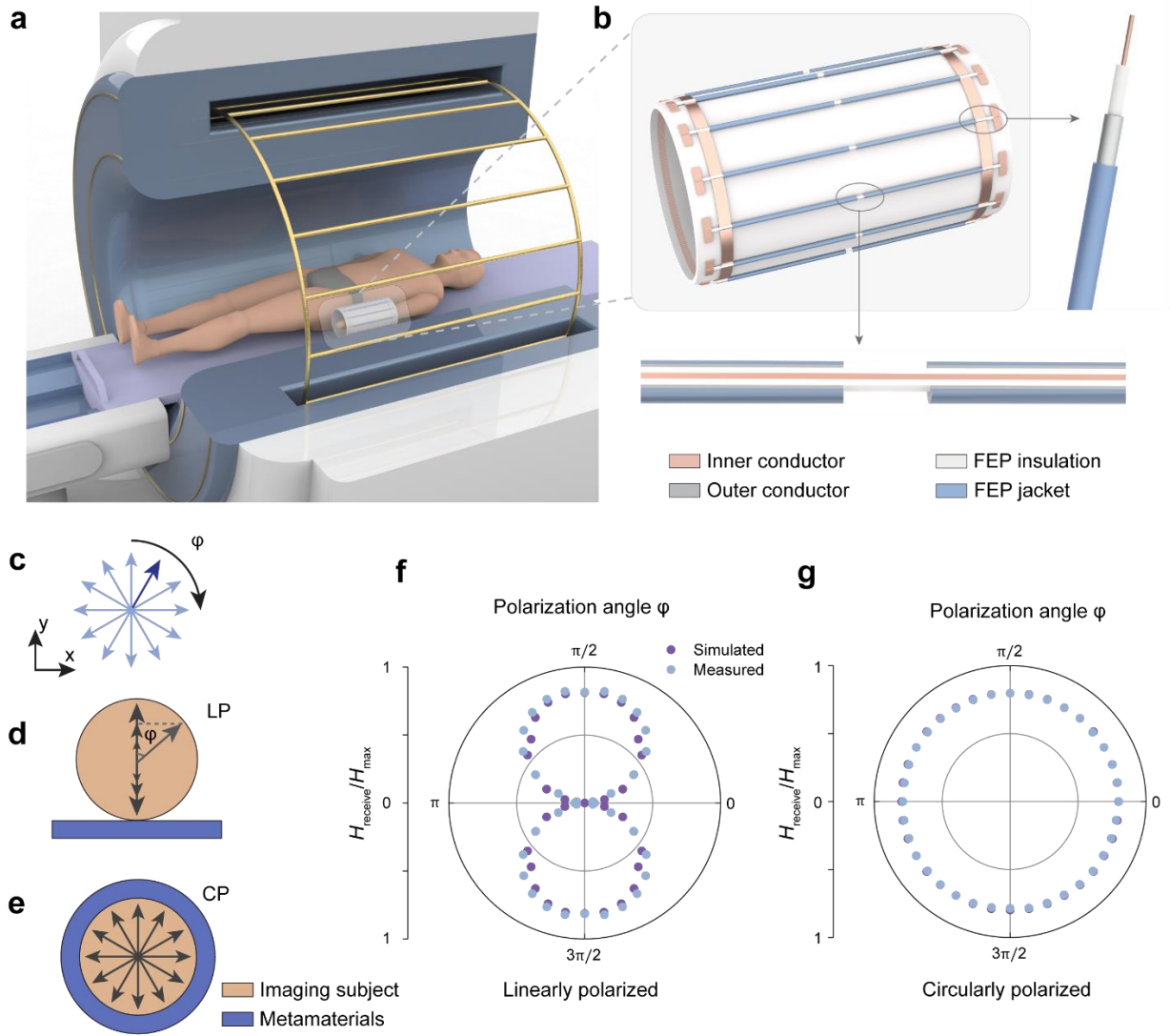

**Figure 1. Circularly polarized metacage for wirelessly enhanced MRI performance. a)** Schematics of the MRI scan with proposed wireless metacage that is form-fitted to the patient's arm. The MRI scanner model was created in SolidWorks and rendered in Keyshot. **b)** Schematics of the metacage comprising 12 independent legs made by coaxial cables with a gap in the outer conductor. The outer conductors of the legs are connected by two end rings at both ends. **c-e)** In the BC, the RF magnetic field is a circularly-polarized (CP) field that rotates in the transverse (x–y) plane, its phase varies with the polarization angle $\varphi$. Traditional linearly polarized (LP) planar metamaterials (**d**) are only sensitive to the component ($H_y$) aligned to their polarization. In contrast, the metacage (**e**), which supports CP resonance, can respond to both components ($H_x$ and $H_y$), making it sensitive to the full CP RF field. **f-g)** Polar plots of the normalized magnetic field $H\_received/H\_max$ obtained from simulations and measurements as a function of the polarization angle, for (**f**) a planar LP metamaterial and (**g**) the CP metacage.

Exploring metamaterials as wireless alternatives offer benefits such as streamlined integration, enhanced design flexibility, and improved patient comfort. These metamaterials can redistribute and amplify the incoming $B_1^-$ magnetic field in the near-field region, tailoring sensitivity specifically within the ROI, and interacting wirelessly with the receive coil to boost the MRI signal. Earlier studies on metamaterials introduced various arrayed configurations for MRI applications, including 'Swiss rolls' [29], split-ring resonators [30,31], spiral coils [32-34], and parallel wire arrays [35-36]. Some of these designs have demonstrated promising MRI signal enhancement, but their use has been limited by constrained experimental setups and the need for manual calibration of MRI protocols and pulse sequences. More recent efforts have aimed to improve clinical compatibility across various MRI vendors and pulse sequences by incorporating passive detuning features and more form-fitting designs, thereby broadening applicability and further enhancing SNR — even approaching that of conventional receive coil arrays [37-40].

Despite significant advancements, many existing metamaterial designs continue to suffer from substantial field inhomogeneity. Signal enhancement is still confined to regions directly adjacent to the metamaterials, while deeper anatomical areas exhibit significantly reduced SNR. This spatial non-uniformity can degrade image contrast, reduce overall image consistency, introduce artifacts, and ultimately compromise diagnostic accuracy—particularly in planar metamaterial designs, where the magnetic field strength decays rapidly along the normal direction of the structure [39,40,53]. The critical limitation of most existing designs is their reliance on linearly polarized resonance modes. In contrast, the $B_1^-$ field in MRI is inherently circularly polarized, comprising two components, $H_x$ and $H_y$, that are spatially and electrically orthogonal [41]. As a result, linearly polarized metamaterials only enhance the component aligned with their polarization, leaving almost half of the incoming RF energy unutilized. This mismatch leads to SNR degradation and severely limits their effectiveness in anatomically complex regions. Circular polarization therefore provides a unique advantage by enabling the metamaterial to couple constructively to both quadrature components of the RF field, thereby fully utilizing the available $B_1^-$ energy. This improved quadrature detection significantly enhances overall imaging performance, offering a theoretical signal sensitivity increase of a factor of $\sqrt{2}$ compared to equivalent optimized linearly polarized designs within the ROI [42]. However, achieving simultaneous strong signal enhancement and volumetric field homogeneity using a circularly polarized metamaterial presents substantial design challenges. This difficulty arises from the complex electromagnetic design and analysis required to engineer phase-coherent resonant modes, the sensitivity of circularly polarized responses to geometric tolerances, and the inherent difficulty of characterizing and validating the resonance behavior of fabricated metamaterials. Volumetric metamaterials have been explored as an alternative and can offer improved field homogeneity; however, such designs are often bulky, difficult to control in terms of resonant mode distribution and inter-element coupling, and tend to deliver suboptimal SNR [43,44]. Some more recent circularly polarized metamaterial [37,38,45] provides a simplified architecture and improved polarization control, yet its reliance on conventional conductors limits geometric scalability and constrains the achievable SNR enhancement.

In this work, we introduce a wireless metamaterial cage, referred to as the "metacage," which features a customized electromagnetic response and achieves substantial SNR enhancement together with exceptional volumetric field homogeneity, distinguishing it from prior approaches. The coaxial unit cells are connected through the shared end ring, ensuring that the induced currents always combine constructively, enabling a simplified yet robust circularly polarized interaction with the RF receive field B1-. To present and validate this unique circularly polarized behavior, we developed and experimentally implemented a visualization framework that reveals the phase-shifted electromagnetic field distribution within the structure. Such methodology provides a generalizable approach for analysis and design of future circularly polarized wireless metamaterial system. Furthermore, the metacage is constructed from off-the-shelf, non-magnetic coaxial cables, which are modified to function as resonant building blocks in an arrayed configuration. Each element inherits the coaxially shielded structure of the original cable, effectively confining the electric field within the structure when on resonance. This feature promotes safer clinical integration and reduces undesired electric field coupling [46,47], ultimately contributing to further SNR improvement. The performance of the metacage is thoroughly characterized through simulation, bench measurement, and MRI validations, demonstrating its ability to provide strong and uniform SNR enhancement across the targeted ROI.

## 2 RESULTS

The metacage is designed as a cylindrical structure with a diameter of 100 mm and a height of 155 mm, comprising 12 identical metamaterial unit cells (legs) symmetrically arranged around the central axis of the cylinder and interconnected through two circular end rings (**Figure 1b**). The structure's resonance frequency is optimized to align with the Larmor frequency in a 3.0 T MRI system (127.7 MHz). Each unit cell is constructed from a selected non-magnetic coaxial cable (9849, Alpha Wire) with a 2.54 mm diameter whose distributed inductance, resistance, and dielectric geometry are compatible with resonance at the 3.0 T Larmor frequency in the targeted imaging volume. The cable comprises four layers from the inside out: the inner conductor (silver-plated copper braid), the fluorinated ethylene propylene (FEP) insulation, outer conductor (silver-plated copper braid shield), and FEP jacket. While commercially available, the cable requires structural modification to operate as a metamaterial element: the outer conductor is interrupted by a 3-mm engineered gap and reconnected at both ends via a copper end-ring, defining the distributed reactance of the unit. The inner conductor remains isolated from neighboring legs but is connected to its own outer conductor through two oppositely oriented PIN diodes (MADP-000235-10720T, MACOM Technology Solutions) (Figure S1), enabling nonlinear passive detuning. The metacage design is scalable to other field strengths through geometric tuning or by selecting coaxial cables of different distributed reactance (Tables S31). A 3D-printed cylindrical scaffold is used to align and secure the metacage, ensuring a stable operating frequency and enhancing robustness for conformity to the anatomical ROI. Traditional metamaterials designed for near-field-based applications like MRI often employ helical or spiral geometries, leveraging their inter-turn distributed capacitance to achieve resonance at the Larmor frequency without requiring larger physical dimensions. However, these structures are highly sensitive to environmental perturbations, resulting in instability and frequency drift [33]. Alternative approaches, such as capacitively-loaded rings using discrete lumped components, introduce added complexity and can degrade the quality factor due to parasitic losses [30]. In contrast, the coaxial cables used in the metacage offer critical electromagnetic advantages, particularly in enabling a compact physical footprint through the distributed shunt capacitance naturally formed between their inner and outer conductors. However, directly implementing coaxial cables as conductive materials for metamaterials fails to produce the desired resonant behavior due to excessive shielding. An intact outer conductor acts as a Faraday cage, isolating the inner conductor and preventing the external RF field from penetrating the structure, effectively eliminating any functional capacitance between the conductors (Figure S2). This limitation is overcome by introducing a small structural gap in the outer conductor, breaking the unwanted shielding effect. This modification enhances coupling to the external field while increasing the cable's effective self-capacitance, making it possible to tune the resonance to the Larmor frequency (127.74 MHz at 3.0 T) without requiring physical dimensions on the order of the RF wavelength (~2.35 m) or the use of additional lumped elements.

The cylindrical symmetry of the metacage is a critical feature, as it aligns naturally with the circularly polarized nature of the MRI receive field (**Figure 1c**). This structural arrangement enables the metacage to interact uniformly with the rotating magnetic field components, facilitating efficient coupling across all polarization angles. To demonstrate this process, we compare the simulated and measured magnetic fields, characterized by reflection coefficients, generated by the metacage and a conventional planar metasurface under rotating excitation polarizations **(Figure S3).** The planar metasurface exhibits strong angular dependence, with nulls appearing in specific directions due to its linearly polarized response (**Figure 1d and 1f**). In contrast, the metacage shows a consistent response across all orientations, confirming its ability to support circularly polarized resonance by simultaneously coupling to both $H_x$ and $H_y$ components (**Figure 1e and 1g**).

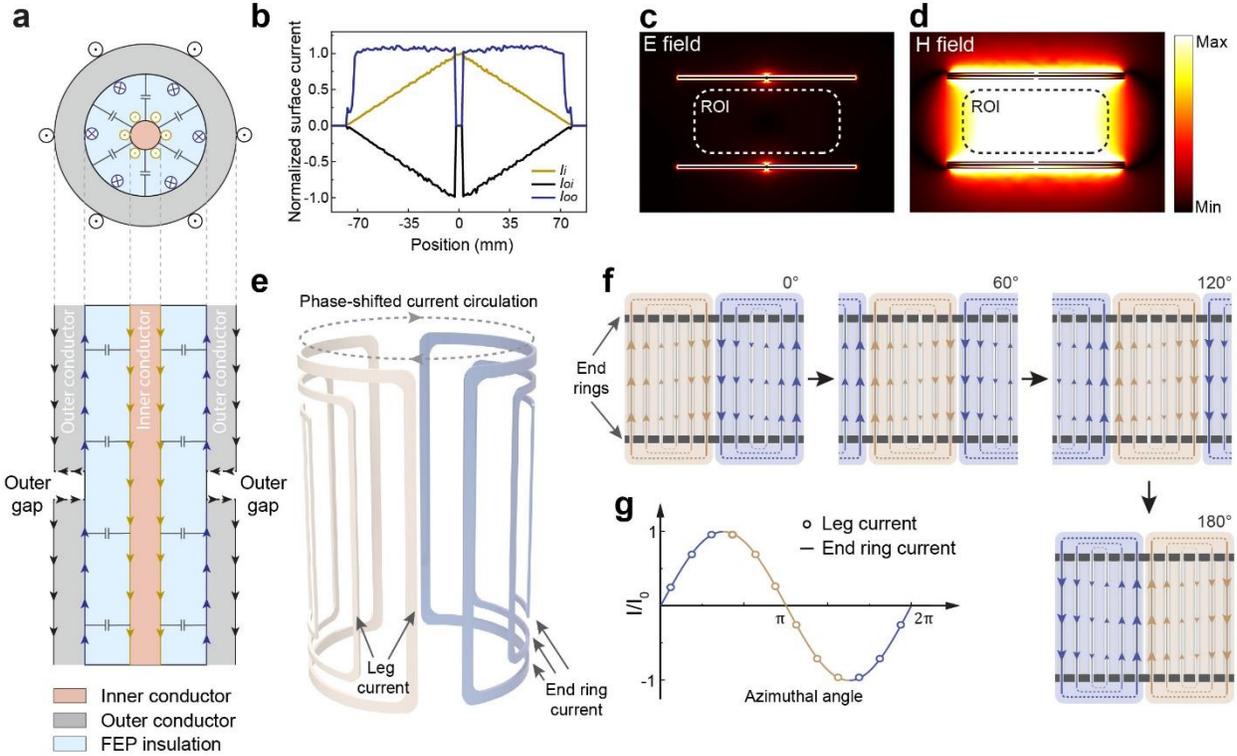

**Figure 2. Symmetric current loops and resonance behavior in the metacage. a)** Electric current distribution within the coaxial cable after introducing the outer gap. **b)** Integrated surface currents distribution along the legs. **c-d)** Simulated electric and magnetic field within the metacage at resonance, indicating the electric field is mostly confined within the cable instead of being exposed in the ROI (**c**), while the magnetic field is uniform and concentrated in the ROI (**d**). **e)** Schematic of the instantaneous three-dimensional current distribution on the metacage. Vertical traces represent the currents flowing along the coaxial legs, while horizontal traces at the top and bottom represent the currents on the end rings. The end-ring currents are slightly offset to illustrate how the N legs form N/2 closed current loops, using the end rings as shared paths. **f)** Unwrapped cylindrical surface showing the phase-shifted current distribution in a 12-leg metacage. The beige and blue shaded regions highlight two groups of current loops and their phase progression along the end-ring periphery. Each group contains three loops, and each loop consists of a pair of legs connected through the end rings. The four panels correspond to four phase instants (0°, 60°, 120°, 180°) as the circularly polarized B1- excitation rotates. **g)** Normalized current magnitude along the end ring at the 90° phase instant. This sinusoidal profile results in a hierarchical current distribution, in which larger loops carry stronger currents.

When the metacage is excited at its resonance frequency by an alternating RF magnetic field, the currents flow predominantly along the surfaces of its conductors due to the skin effect. As shown in **Figure 2a** and Figure S4, three distinct but related currents are distributed within the coaxial cable in each unit cell: the current along the surface of the inner conductor $I_i$, the induced current along the inner surface of the outer conductor $I_{oi}$, and the current along the outer surface of the outer conductor $I_{oo}$. This unique current distribution is attributed to the presence of the gaps in the outer conductors of each leg. The gap disrupts the shielding conductor layer that would otherwise isolate the inner conductor, effectively mitigating the Faraday cage effect and allowing the RF field to interact with the coaxial cable's interior, facilitating current exchange between the inner and outer surfaces of the outer conductor. **Figure 2b** presents the simulated average current magnitudes along a metamaterial unit cell. The current $I_i$ flows along the inner conductor, exhibiting a symmetric distribution that peaks at the outer gap and reduces linearly toward zero at both ends. Simultaneously, $I_{oi}$ flows in the opposite direction on the inner surface of the outer conductor, following the same profile and also peaks at the outer gap. In regions away from the gap, $I_i$ and $I_{oi}$ maintain equal amplitudes but opposite directions, effectively canceling each other's contribution to the inductive magnetic field. In contrast, $I_{oo}$ remains nearly uniform along the cable length and serves as the primary source of the induced magnetic field. This uniform current distribution helps minimize spatial variation in the generated magnetic field, offering a distinct advantage over traditional parallel conductors-based, dipole-like metamaterial configurations, which typically exhibit sinusoidal current profiles and increased field non-uniformity [35].

Another benefit of allowing the interior of the coaxial cable to participate in resonance through the outer gap is the introduction of the structural capacitance between the conductor layers, similar to the shunt capacitance observed in coaxial cables when used as transmission lines. The metacage is passively detuned during the transmit phase and becomes active during the receive phase, ensuring that it does not interact with the high-power B1+ field. When resonant in the receive phase, the coaxial geometry determines where the dominant capacitive energy of the mode is stored, leading to electric fields that are primarily concentrated within the dielectric between the inner and outer conductors. This localization arises directly from the coaxial geometry, where the resonant electric energy is stored internally rather than in the surrounding sample. As a result, stray electric fields in the tissue-facing region are substantially reduced, while the magnetic resonance remains efficiently supported to generate a strong induced B1- field within the ROI (Figure 2c-d). By limiting capacitive coupling to adjacent tissue and confining most of the resonant electric-field interaction within the cable structure, the metacage minimizes undesired electric-field exposure during the receive phase and supports safer, lower-noise MRI operation.

However, while the introduction of outer gaps can enhance coupling between the RF field and the metamaterial, increasing the number of gaps can weaken the overall induced current in the cable. As the number of gaps increases, the current is divided into more segments, diminishing the magnitude of the current primarily responsible for generating the induced magnetic field, potentially reducing the overall efficiency (Figure S5). To maintain an optimal balance between coupling efficiency and current strength, the metacage employs a single structural gap in each metamaterial unit cell.

As mentioned, the uniform distribution of $I_{oo}$ along each individual coaxial cable contributes to localized field uniformity, but it is the collective current distribution across the entire cylindrical metacage that fundamentally enables the high level of magnetic field homogeneity required for quality MRI imaging, as exhibited in **Figure 2d**. Specifically, under the working resonance mode, the metacage with N legs can be conceptualized as containing N/2 closed loops, each formed by two legs interconnected through the two end rings serving as shared conductive paths. As shown in **Figure 2e** and Figure S6, at any given moment, these loops are naturally divided into two groups, with currents of equal magnitude flowing in opposite directions. This symmetrical configuration, facilitated by the phase-shifted current circulation around the end rings, enables these current loops to dynamically rotate in response to the external circularly polarized $B_1^-$ field, allowing them to continuously align perpendicularly with the polarization of the $B_1^-$ field (Figure S7), thereby maximizing the RF field interaction.

The currents generated on each leg are sequentially phase-shifted around the end rings' periphery, to further illustrate how the current distribution evolves with the phase of the RF magnetic field, **Figure 2f** presents an expanded, unwrapped view of a metacage model with 12 legs with phase progression of currents at 0°, 60°, 120°, and 180°. For a metacage with N legs, the phase difference between the currents in neighboring elements is 360°/N, resulting in a sinusoidal distribution of current amplitudes across the legs around the end rings' periphery, even though each individual leg carries a relatively uniform current $I_{oo}$ along its length at any given moment (**Figure 2g,** Figure S6). At each phase, the legs can be grouped into two current sets flowing in opposite directions, forming pairs of nested, symmetric current loops. These loops generate magnetic fields that constructively interfere within the metacage, reinforcing the net magnetic field. This configuration ensures that each set of current loop aligns perpendicular to the external magnetic field, enhancing coupling efficiency. Moreover, the hierarchical distribution of current—where larger loops carry stronger currents—naturally enables the magnetic field generated by the metacage to be tailored into a more uniform shape, minimizing spatial variation and ensuring consistent field strength across the ROI [48].

Effective integration of the metamaterial into MRI systems requires precise control over its operating frequency to ensure seamless interaction with the BC and compatibility with standard pulse sequences. Achieving resonance at the Larmor frequency is essential for sustaining the operating mode that supports homogeneous magnetic field generation and optimal signal enhancement. Furthermore, the Larmor frequency can vary across MRI systems, even among 3.0 T platforms from different vendors, again underscoring the importance of design flexibility. The ability to tune the resonance frequency through structural parameters not only facilitates broader clinical applicability but also accommodates variations in patient anatomy. By combining frequency tunability with geometric adaptability, the metacage can be optimized to achieve the best trade-off between magnetic field strength and homogeneity across a wide range of clinical applications.

The resonance mechanism of the metacage is governed by the L-C circuit resonance, where the resonance frequency is given by $f = 1/(2\pi\sqrt{LC})$, with L and C representing the inductance and capacitance of the metamaterial, respectively. In the metacage, the substantial structural capacitance of the coaxial cable serves as the primary source of capacitance, while the inductance primarily arises from the self-inductance and mutual inductance of the metamaterial elements—both of which are strongly dependent on the length and relative positioning of the coaxial cables. Thus, tuning the resonance frequency can be achieved through structural modifications to the metacage. Specifically, key design parameters such as the diameter (D), length (L), and the number of legs (N) directly influence the frequency response (**Figure 3a**), allowing for precise adjustment to match the required Larmor frequency for MRI systems. The metacage can be geometrically tuned to the target resonance frequency of 127.7 MHz (**Figure 3b** and Figure S8), as confirmed by simulations. The plots illustrate a clear trade-off between the design parameters: a reduction in leg length L can compensate for either an increased number of legs N or a larger diameter D, thereby preserving the desired resonance frequency.

To systematically explore the trade-offs among the key design parameters, twelve different metacage configurations are evaluated to compare magnetic field strength, coverage, and field homogeneity. In the first set of simulations, the diameter D is fixed, while the number of legs N varies. In the second set, the number of legs N is held constant while the diameter D changes. **Figures 3c and 3d** visualize how the induced magnetic-field pattern evolves when the geometric parameters D, L, and N are varied under identical excitation conditions, illustrating the general trends that guide practical parameter selection. One-dimensional plot of the simulated magnetic field along the central axis demonstrates that increasing N enhances magnetic field concentration but reduces coverage along the longitudinal direction (**Figure 3e**), while decreasing the diameter D increases field strength and concentration, though at the cost of reduced spatial coverage, potentially limiting applicability for larger anatomical regions (**Figure 3f**). Moreover, since clinical MRI primarily produces two-dimensional images, the metacage design must deliver both strong and uniform magnetic fields across a representative imaging plane. To assess this, we evaluate a rectangular region (60 mm × 120 mm) in the coronal plane and a circular region (80 mm diameter) in the axial plane, approximating the anatomic dimensions of a human wrist or arm segment. The mean magnetic field strength and corresponding homogeneity are evaluated for all twelve configurations (**Figs. 3g and 3h,** Figure S9). Based on the results, the metacage with 12 legs and a 100 mm diameter is selected for further MRI validation due to its optimal balance between field strength, coverage, and field uniformity.

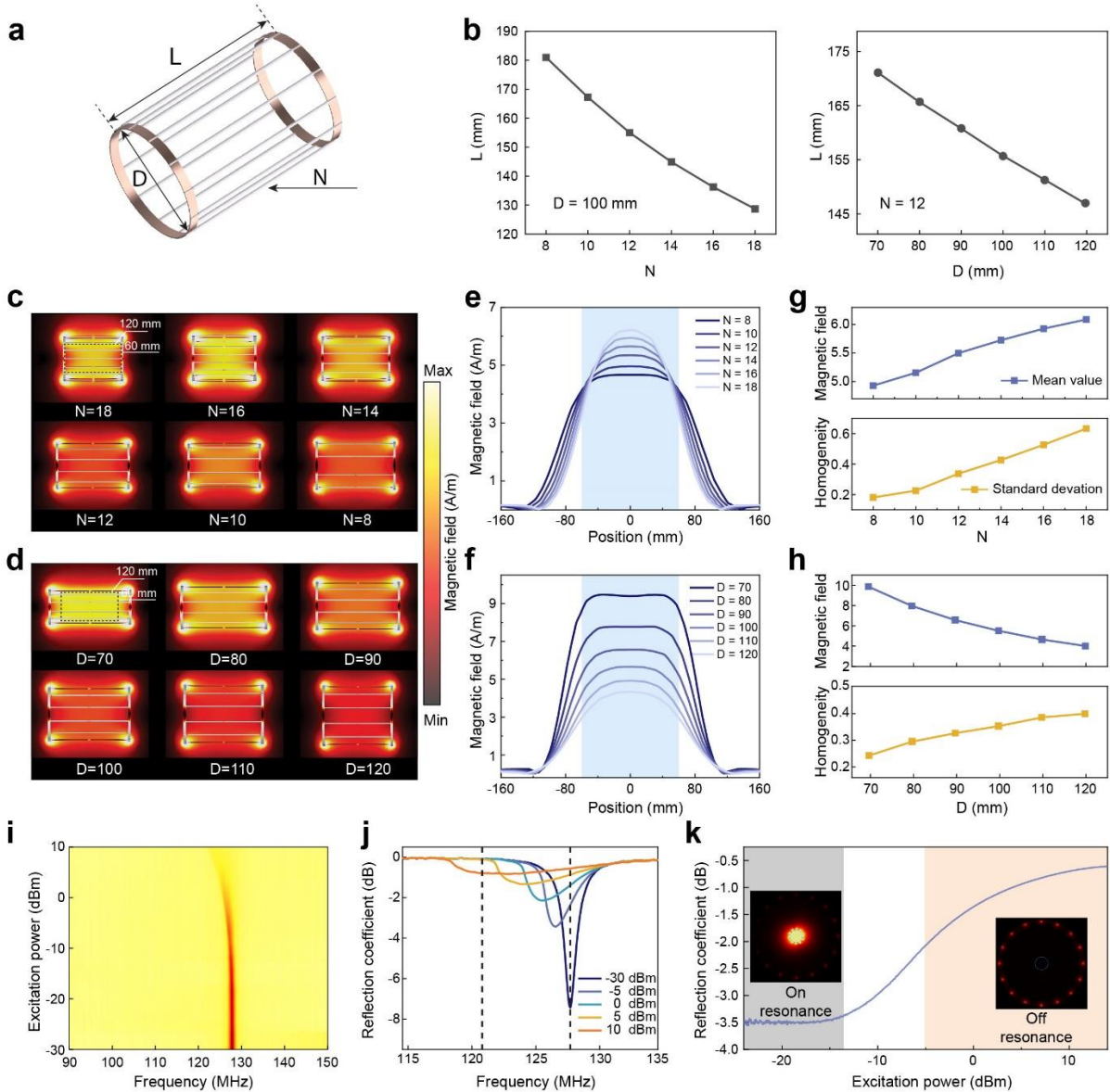

**Figure 3. Magnetic field shaping via geometric modulation and nonlinear response of the metacage. a)** Tunable geometry parameters of the metacage, including the length (L), diameter (D) and number of legs (N). **b)** To maintain alignment with the Larmor frequency, increasing N at a fixed D = 100 mm, or increasing D with a fixed N = 12, requires shorter L. **c-d)** The simulated magnetic field distribution within the metacages with different N from 8 to 18 (D=100 mm) and D from 70 to 120 mm (N=12), while maintaining resonance at 127.7 MHz by adjusting L. Magnetic-field magnitude is reported in A/m. All simulations were performed using a fixed total excitation power of 1 W applied to the birdcage feed ports. No phantom was placed inside the metacage for these simulations. **e-f)** The magnetic field plotted along the central axis of the metacages. **g-h)** The simulated magnetic field (mean value) and homogeneity (standard deviation) in a 2D area (60 * 120 mm) on the coronal planes in the metacages. **i-j)** Measured reflection coefficient at different excitation power from -30 to 10 dBm. **k)** Measured reflection coefficient as a function of excitation power at 127.7 MHz. Insets: Simulated magnetic-field distributions of the metacage placed inside the body coil, shown for the reception and transmission states to illustrate the corresponding resonant and suppressed modes.

To ensure compatibility with MRI systems, metamaterials must remain electromagnetically inactive during transmission while enhancing signal reception. Without this adaptive behavior, the structure would interfere with high-power transmit pulses, making it unsuitable for standard imaging sequences. A common solution is to integrate nonlinear components—such as diodes or varactors—that respond differently to the power level during the transmission ($kW$) and receive ($\mu W$) phases [49]. In this study, we integrate a pair of non-magnetic PIN diodes at the ends of each inner conductor to connect them to their respective outer conductors (Figure S1). This configuration leverages the strong electric field between the two conductors, ensuring that the voltage difference during high-power transmission can forward-bias the diodes. As a result, the metacage detunes passively during the transmission phase and reverts to its resonant state during reception.

The self-adaptive response of the metacage is evaluated by measuring its reflection spectra across a range of excitation powers using a vector network analyzer and an inductive loop antenna. As the excitation power increased from -30 dBm to 10 dBm (**Figs. 3i and 3j**), the resonance frequency shifted downward from the Larmor frequency, accompanied by a reduction in oscillation amplitude. While this power range does not reach the levels used in actual MRI transmission, it demonstrates a clear nonlinear trend consistent with diode rectification: increasing RF power induces a self-bias voltage across the diode, altering the structural capacitance and weakening the resonance at 127.7 MHz (**Figure 3k**) [50]. This behavior supports the expected outcome at higher power levels, where the diodes are anticipated to remain forward-biased throughout transmission, effectively creating a short between the inner and outer conductors and detuning the metacage. This behavior is further validated in a 3.0 T MRI environment using a high-pass BC model in CST Microwave Studio. Reception was simulated by directly exciting the structure in its resonant state, while transmission was modeled by shorting the inner and outer conductors to represent forward-biased diodes. As shown in **Figure 3k**, the metacage significantly enhances the local magnetic field—by a factor of ~342—during reception, while remaining electromagnetically inactive during transmission. Its circularly polarized response ensures uniform coupling to the rotating field, confirming effective passive detuning under realistic MRI conditions.

To experimentally validate the metacage's passive detuning behavior in a real MRI environment, we analyze how the SNR responds to a range of prescribed flip angles (FAs) in a uniform mineral oil phantom using gradient echo imaging. The trend of SNR versus nominal FA provides an indirect assessment of whether the transmit field $B_1^+$ was affected by the presence of the metacage [51]. If passive detuning is effective, the transmit field—and thus the actual FA—should remain unchanged. As shown in **Figure 4a** and Figure S10, the metacage significantly increases SNR, but the trend as a function of FA closely follows that of the body coil alone. This consistency suggests that the transmit field remains unaltered, confirming the intended nonlinear behavior of the metacage during the transmission phase.

Additionally, evaluating the SAR is essential to ensure patient safety, as excessive RF energy can lead to tissue heating. To assess the safety of the metacage in a realistic scenario, a human wrist model was used for SAR simulations. As shown in the simulated SAR map in Figure 4b, a modest change in the peak local SAR was observed, increasing from 0.388 W/kg in the baseline configuration to 0.428 W/kg when the metacage was present. This variation is consistent with the localized electric-field redistribution around passive conductive structures and remains well below the IEC 60601-2-33 limits for normal-mode operation (Figure S11). These advantages can also be attributed to the nonlinear and selective response of the metacage, which ensures compatibility with the BC and maintains safety without introducing unpredictable effects during standard MRI operation. In addition to SAR simulations, proton resonance frequency shift (PRFS) thermometry was performed to assess RF-induced temperature changes under high-power heating conditions. The metacage exhibited phase-drift behavior comparable to that of a commercial extremity coil, indicating that no additional heating attributable to the metacage was observed under the tested conditions (Figure S12).

The SNR enhancement achieved by the metacage is experimentally validated using a clinical 3.0 T MRI scanner (Philips Healthcare) and a mineral oil phantom. Gradient echo imaging sequence is used for the scans. Figures 4c to 4e show the experimental setup, where a uniform phantom with a diameter of 100 mm is directly placed inside the metacage. During image acquisition, the BC is used for both signal transmission and reception.

For comparison, imaging is also performed using a commercially available 16-channel extremity coil (Philips Healthcare) and a PCB-based metacage of the same size that lacked the coaxial cable structure. Figures 4f to 4h show the corresponding SNR maps from axial and coronal slices for each configuration. Compared to the 16-channel coil, the coaxial cable-based metacage demonstrates nearly uniform SNR distribution in both axial and coronal planes. To quantitatively assess the SNR enhancement, we evaluate the SNR at the central region of the phantom, as indicated by the dashed line in Figure 4f. All measured SNR values are normalized to the baseline case acquired using the BC alone (Figs. 4i to 4j). When compared with the simulated SNR, which reflects the theoretical behavior, the measured results show the same trend, with only minor magnitude differences expected between practical measurements and idealized simulations (Supporting Note S4, Figure S13). In the axial plane, the coaxial metacage demonstrates exceptional homogeneity, with SNR variations within 12.07 % of the peak value, compared to 54.83 % for the 16-channel extremity coil. The minimum SNR enhancement ratio achieved by the coaxial metacage is 30.64, significantly higher than the 15.76 and 19.16 observed for the 16-channel coil and PCB-based metacage, corresponding to enhancements of 1.94-fold and 1.60-fold, respectively. In the coronal plane, the coaxial metacage also delivers a highly concentrated SNR distribution, with improvements of up to 2.24-fold and 1.41-fold over the extremity coil and PCB-based metacage. It is worth noting that although the PCB-based metacage also demonstrates comparable homogeneity in both axial and coronal planes, it is consistently outperformed by the coaxial design. This performance gap is primarily attributed to the absence of effective electric field confinement in the PCB configuration, which leads to greater capacitive losses, increased noise, and consequently, reduced SNR. These results confirm that the proposed coaxial cable-based metacage significantly enhances both SNR and imaging homogeneity in MRI applications, providing a robust alternative to conventional receive coil arrays.

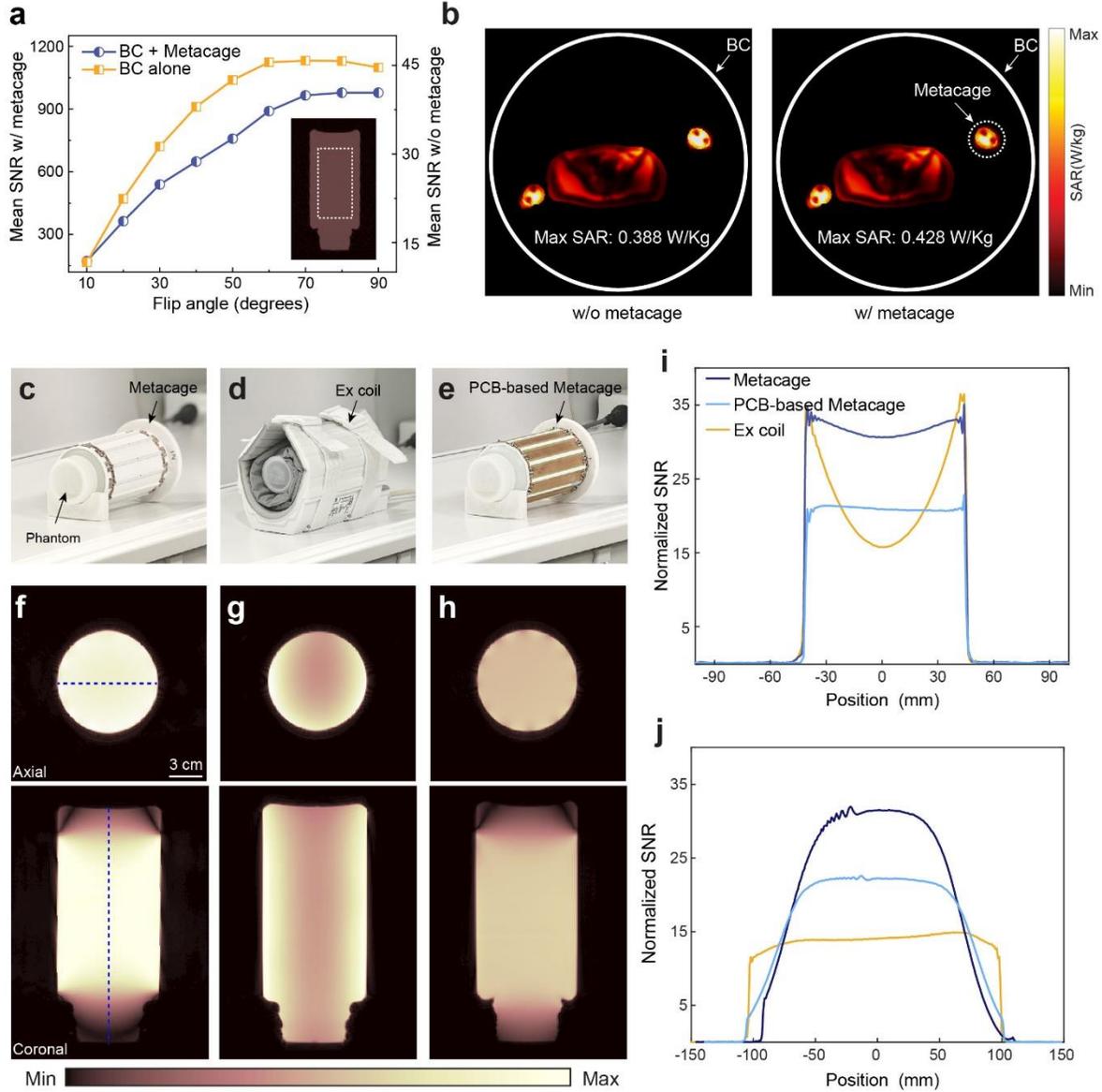

**Figure 4. Metacage-enhanced homogeneous SNR map of a phantom. a)** Mean SNR (evaluated in the dashed rectangular region) as a function of the flip angle with and without the metacage using gradient echo imaging. **b)** Simulated SAR maps (10 g, normalized to 1 W of accepted power) acquired with and without metacage in the axial plane. **c-e)** MRI imaging setup of the proposed metacage (**c**), extremity coil (Ex coil) (**d**), and PCB-based metacage (**e**) with a mineral oil phantom. **f-h)** Axial and coronal SNR maps acquired with the metacage (**f**), extremity coil (**g**), and PCB-based metacage (**h**). **i-j)** SNR profiles along the dashed line in axial planes (**i**) and coronal planes (**j**).

In clinical practice, MRI employs a wide range of pulse sequences tailored to specific anatomical targets and diagnostic purposes. Therefore, it is essential that the metacage is compatible with routine clinical imaging protocols. To evaluate this compatibility, five commonly used clinical pulse sequences were tested: gradient echo (GRE), proton density-weighted turbo spin echo (PDw TSE), proton density-weighted turbo spin echo with fat saturation (PDw TSE (SPIR)), T1-weighted turbo spin echo (T1w TSE), and T2-weighted turbo spin echo (T2w TSE). These sequences are applied to image the same porcine leg sample to simulate realistic clinical conditions. For each sequence, imaging is performed under three configurations: (1) 16-channel extremity coil with BC, (2) metacage with BC, and (3) BC only. The resulting MRI images are depicted in **Figure 5a**, with the corresponding experimental setup provided in Figure S14. Additional MRI scans were conducted on a smaller biological sample, a grapefruit (Figure S15), demonstrating the capability of the metacage for imaging multiple anatomic subjects. Each of these sequences is optimized to highlight specific tissue contrasts, assisting clinicians in accurately assessing physiological and pathological changes.

The metacage is compatible with all tested sequences, and it consistently demonstrates significant SNR enhancement without introducing artifacts. The images exhibit high uniformity and improved contrast compared to those acquired using the BC alone. Compared to the commercially available 16-channel extremity coil, the metacage provides focused and substantial enhancement within its targeted area. To quantitatively evaluate SNR, we select three regions of interest (ROIs): one bone marrow segment and two muscle segments. Across all pulse sequences (Figure 5b), the metacage produces SNR values in both bone marrow and muscle tissues that are comparable to, or even exceed, those achieved with the 16-channel extremity coil, and both cases yield SNR substantially higher than those obtained using the BC alone. Notably, each clinical pulse sequence imposes distinct signal-weighting conditions—such as TR/TE timing, flip angle, receive bandwidth, echo-train length, and fat-suppression modules—which inherently produce different baseline SNR levels even when the same receive coil is used. For example, SPIR and long-TE T2w-TSE sequences naturally yield lower absolute signal, compressing the apparent enhancement ratio, while tissue-specific relaxation properties (T1, T2, and proton density) further contribute to the differences observed between muscle and bone marrow. It is also important to recognize that the metacage operates wirelessly by modifying the B1- receive sensitivity of the BC, whereas the extremity coil functions as an independent multi-channel array that directly receives the MR signals. Although both configurations use the BC for transmission, they represent fundamentally different receive mechanisms. Therefore, the comparison between the metacage with BC and the extremity coil with BC is best interpreted qualitatively. Under this qualitative assessment, the metacage provides higher SNR in most sequences, demonstrating robust performance and strong generalizability across diverse clinical imaging conditions. A key advantage of the metacage is its entirely wireless operation, seamlessly coupling with the BC without the need for complex wiring or additional hardware. This wireless configuration simplifies the experimental setup while significantly reducing manufacturing costs. The metacage's circularly polarized response and well-defined tuning mechanism allow for convenient and cost-effective customization of the geometry, enabling straightforward adaptation to various anatomical regions in multiple MRI systems. This flexibility makes the metacage a versatile and scalable solution for clinical MRI enhancement, offering a practical and efficient approach to improving image quality across diverse diagnostic applications.

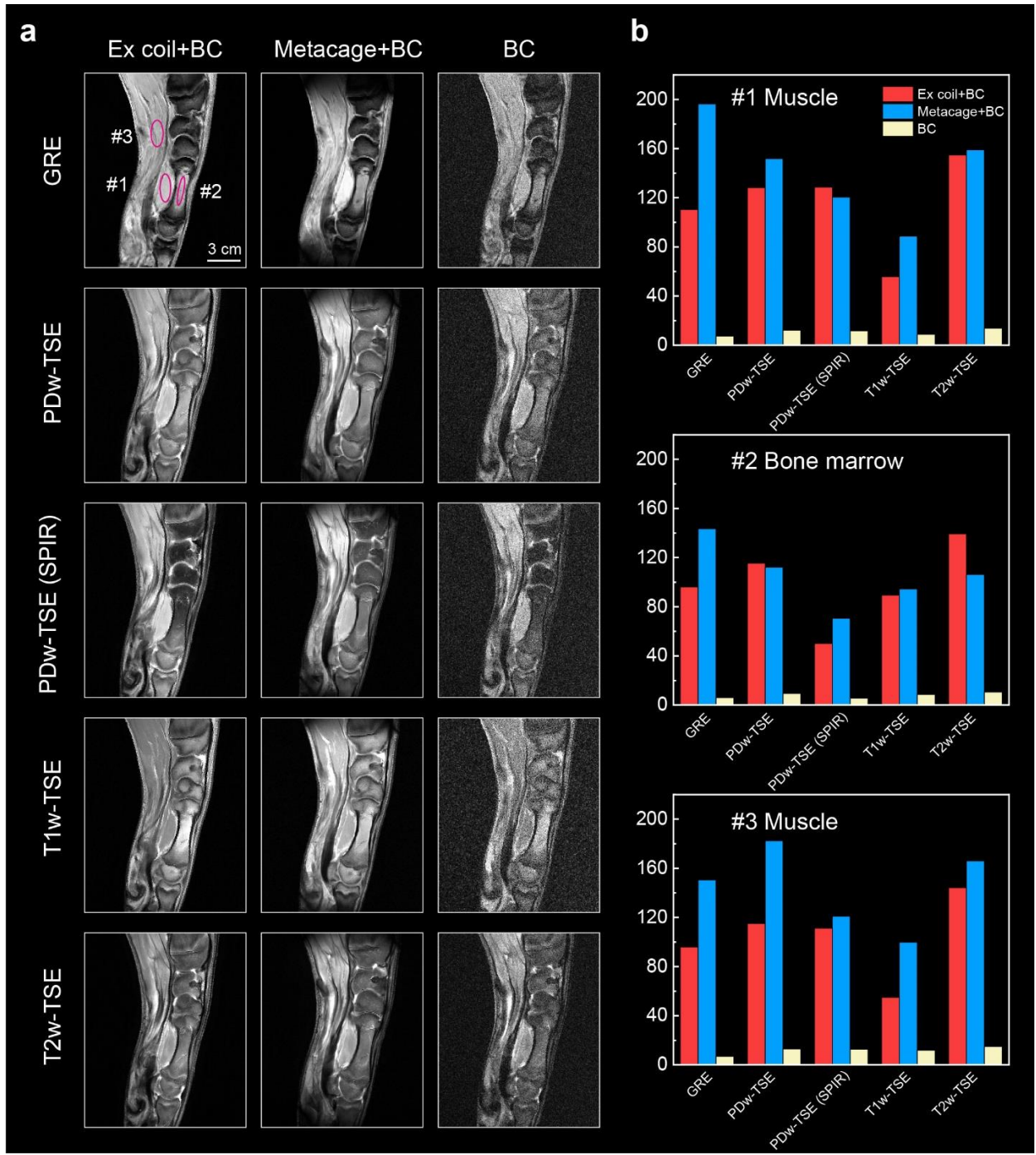

**Figure 5. MRI validations with the ex vivo porcine leg sample. a)** MRI images of the porcine leg acquired with the extremity coil (Ex coil) and BC, metacage and BC (Metacage+BC), and BC only, using

5 clinical MRI pulse sequences: GRE, PDw TSE, PDw TSE (SPIR), T1w TSE, and T2w TSE. **b)** Quantitative evaluation of the SNR for different regions #1, #2, and #3 as outlined.

## 3 DISCUSSION

In this study, we present a wireless metamaterial cage for enhanced SNR in MRI, specifically tailored for extremity imaging on 3.0 T MRI systems. The metacage is validated through ex vivo experiments, demonstrating compatibility with standard clinical pulse sequences and delivering comparable or superior SNR when compared to the most advanced commercial extremity receive coils. The metacage offers several critical advantages. Constructed from commercially available coaxial cables, the metamaterial is inherently low-cost, lightweight, highly customizable, and optimized for patient comfort. A key feature of the metacage is its circular polarization, enabled by its cylindrical symmetry and sequentially phase-shifted current distribution, which allows simultaneous coupling to both $H_x$ and $H_y$ components of the MRI receive field $B_1^-$, regardless of the polarization angle. This quadrature detection enhances coupling efficiency, leading to improved SNR compared to linearly polarized designs. Notably, the metacage achieves exceptional magnetic field uniformity across the ROI when compared to the extremity coil, minimizing signal variation and ensuring consistent imaging quality. This uniform field distribution is a direct result of the carefully engineered metamaterial design. Moreover, the metacage is compatible with parallel imaging techniques when used in combination with a coil array, enabling accelerated image acquisition. This capability was demonstrated using a Philips spine coil — chosen for its widespread clinical use and integration within the patient bed, which ensured a clean and simple setup — at acceleration factors of 2, 3, and 4 (Figure S16). Importantly, the metacage does not introduce image artifacts and enhances SNR homogeneity across varying distances from the imaging subject, further supporting its broad clinical utility. Compared to existing volumetric metamaterials, another key innovation of the metacage is its coaxial cable-based construction, which naturally confines electric fields and reduces parasitic capacitive coupling to surrounding tissues. This design improves SNR and enhances patient safety by reducing local SAR. The introduction of a structural gap in the outer conductor of each unit cell allows precise resonance matching to the Larmor frequency without requiring additional lumped elements, maintaining both simplicity and efficiency. In addition, the integration of PIN diodes provides a passive detuning mechanism that ensures the metacage remains electromagnetically inactive during the transmission phase and active only during reception, thereby avoiding any alteration of the transmit field or tissue relaxation dynamics. This nonlinear behavior is essential for compatibility with standard clinical imaging protocols and is validated through simulation, bench measurements, and MRI scans**.** Importantly, the metacage is fully compatible with contrast-enhanced MRI. The longitudinal (T1) and transverse (T2) relaxation times, as well as the relaxivity-driven effects of paramagnetic agents such as gadolinium chelates or iron-oxide particles, are governed by microscopic spin-lattice and spin-spin interactions within the local molecular environment. The observed image contrast depends on the characteristics of the transmission excitation pulses applied in the imaging sequence. For both situations, the metacage neither alters the local magnetic microenvironment of the tissue nor participates in the transmission; it cannot modify these relaxation mechanisms or contrast-agent behavior. Its influence is limited to improving the receive sensitivity, analogous to conventional receive-only coils.

Experimental MRI imaging results using phantoms and ex vivo tissue demonstrate the metacage's ability to deliver uniform, high-SNR images across multiple clinical sequences. Compared to a commercial 16-channel extremity coil, the metacage provides comparable or superior performance within its targeted imaging region, despite operating wirelessly without additional cabling and circuits. This efficiency, combined with its ease of fabrication and adaptability, makes the metacage a cost-effective, patient-specific solution for clinical MRI enhancement, especially in applications requiring compact or anatomically conformal coil designs. Nevertheless, while the metacage design supports circular polarization and high field uniformity, its configuration is currently optimized for small extremity imaging. Extending this cylindrical design to larger anatomical regions such as the knee or head is a promising direction, although this will require scaling strategies, potentially involving modular designs or nested structures. These strategies must carefully balance the trade-off between field coverage and SNR, as increasing the size or complexity of the metamaterial may compromise signal enhancement. Additionally, while passive detuning has been shown to be effective, future work could explore more precise control using active components or smart materials for dynamic tuning across variable field strengths and imaging conditions. Extending the metacage concept beyond 3.0 T also represents a promising direction for future development. Taking the inherent advantage of the geometry detuning property of the metacage, it can be transitioned from 3 T (127.7 MHz) to UHF MRI (e.g., 7 T at 298 MHz) or Lower field (e.g., 1.5 T at 63.8 MHz) by further adjusting its leg length L, diameter D, or the number of legs N and will not influence its circularly polarized or detuning capabilities of the metacage. These factors suggest that, with appropriate geometric refinement, the metacage concept is extendable across a wide range of MRI platforms. Future work will also include structured in vivo studies to evaluate SNR behavior, anatomical variability, tissue loading, and motion under routine clinical conditions. These studies will build upon the engineering validation reported here and establish the metacage's performance and safety in realistic biological environments.

In summary, the metacage introduces a new class of wireless, circularly polarized metamaterials that combine strong SNR enhancement, field uniformity, safety, and clinical practicality. This work opens the door for broader integration of metamaterial-based receiving structures into mainstream MRI systems and provides a foundation for future innovations in compact, customizable, and patient-centric coil design.

## 4 MATERIALS AND METHODS

**Metamaterial construction.** Each leg of the metamaterial cage was constructed using a 155 mm-long coaxial cable segment. A 3 mm section of the outer conductor was removed at the center of each leg, creating a structural outer gap. The outer conductor at each end of the coaxial cable was exposed and securely welded to a 5 mm-wide copper tape end ring. To enable passive detuning, a pair of oppositely oriented PIN diodes was connected between the inner and outer conductors at each end. The non-magnetic coaxial cable used in the metacage comprises four layers: an inner conductor, an insulating dielectric, an outer conductor, and an external jacket, with diameters of 0.48 mm, 1.42 mm, 1.93 mm, and 2.54 mm, respectively. For comparison, the PCB-based metacage was fabricated using a printed circuit board (PCB) prototype machine (ProtoMat S64, LPKF). The material of the PCB dielectric was FR-4.

**Numerical simulation.** Electromagnetic simulations of the magnetic field, electric field, and current distribution were performed using the frequency domain solver in CST Microwave Studio Suite 2021. The metacage was positioned at the isocenter of a constructed high-pass BC model, which consisted of 16 legs, each 800 mm long. The BC was driven by two discrete ports with a 90° phase difference, generating a circularly polarized magnetic field.

SAR simulations were conducted using the time domain solver in CST, with a human voxel model "Gustav" from the CST voxel family. SAR values were calculated using the MRI toolbox in CST during post-processing and were normalized to 1 W of accepted power. PIN diodes were modeled as 1.2-pF capacitors in the receive state and as short-circuited paths in the transmit state, consistent with established approximations used in metamaterial MRI studies.

Electromagnetic simulations were performed in CST Microwave Studio to analyze the angular-relied response of the metamaterial structures. A linearly polarized RF magnetic field at 127.7 MHz was used as excitation, and the metamaterial was rotated from 0° to 360° in 10° increments. At each angle, the induced magnetic field was sampled by an H-field monitor positioned 15 cm above the structure. The maximum value $|H_{max}|$ was obtained when the metamaterial directly faced the excitation field. The normalized ratio $|H_{received}|/|H_{max}|$ was computed and used in Figures 1f-g.

Field-distribution analysis was conducted by driving the BC with a fixed total input power of 1 W in CST Microwave Studio. The metacage was simulated without phantoms present to isolate the influence of geometric parameters (D, L, N) on the intrinsic magnetic-field distribution.

**Bench measurement.** Reflection spectra measurements were conducted using a vector network analyzer (VNA, P5020B, Keysight Inc.) connected to an inductive loop antenna for excitation. The angular dependence of the metamaterial's magnetic response was experimentally characterized using the bench measurement setup illustrated in Figure S3. A single-loop antenna was used for both excitation and reflection measurement and was rotated around the stationary metamaterial in 10° increments from 0° to 360°. The induced magnetic response at each orientation was extracted from the measured reflection spectra and normalized following the same procedure used in the simulations.

**MRI Validation.** Details regarding the MRI scan parameters can be found in Table S3 in the Supporting Information.

**Data availability**

The data that support the findings of this study are available from the corresponding author upon reasonable request.

**Acknowledgements**

This research was supported by the Rajen Kilachand Fund for Integrated Life Science and Engineering. The authors thank the Boston University Photonics Center for technical support. The porcine leg samples used in this work were obtained from a local butcher shop and were commercially available food products. Therefore, the experiments did not involve live animals or human subjects and were not subject to ethical approval by our Institutional Review Board. We confirm that no special ethical approval was required for the use of these materials.

**Author contribution**

Y. Liu and X. Zhu. contributed equally to this work. Y. Liu, and X. Zhang conceived the study. Y. Liu, X. Zhu, K. Wu and X. Zhang designed and constructed the metamaterial. Y. Liu, X. Zhu, K. Wu and X. Zhang designed and conducted the bench measurements. Y. Liu, X. Zhu, K. Wu and X. Zhang conducted the MRI scans. All authors participated in discussing the results. Y. Liu, X. Zhu, S. W. Anderson, and X. Zhang wrote the manuscript.

**Competing interests**

The authors have filed patent applications on the work described herein, application No.: 16/002,458, 16/443,126, and 17/065,812. Applicant: Trustees of Boston University. Inventors: Xin Zhang, Stephan Anderson, Guangwu Duan, and Xiaoguang Zhao. Status: Active.

**Additional information**

Additional information is available for this work.